\documentclass[notitlepage]{revtex4-1}
\usepackage[utf8]{inputenc}
\usepackage[T1]{fontenc}
\usepackage{amsmath}
\usepackage{amsthm}
\usepackage{graphics}
\usepackage{graphicx}
\usepackage{subfigure}
\usepackage{amssymb}
\usepackage{graphics}
\usepackage{graphicx}
\usepackage{epstopdf}
\usepackage{color}
\usepackage{subfigure}
\usepackage{url}
\usepackage{float}
\usepackage{soul}
\usepackage{xfrac}
\usepackage{nicefrac}

\def\x{{\bf x}}

\def\p{\textbf{p}}
\def\q{\textbf{q}}

\def\Vp{V_{\rm p}}

\def\Mp{M_{\rm p}}
\def\Rp{a_{\rm p}}

\newcommand\dertt[1]{ \frac{\partial{ #1}}{\partial t} }

\newcommand\grad{\mbox{${\bf \nabla}$}}

\def\DEL#1{{\textcolor{green}{\textst{#1}}}}    % suggested deletions
\def\DEL2#1{{\textcolor{green}{{#1}}}}    % suggested deletions
\begin{document}
%%%%%%%%%%%%%%%%%%%%%%%%%%%%%%%%%%%%%%%%%%%%%
\title{Interaction between active particles and quantum vortices leading to Kelvin wave generation}
\author{Umberto Giuriato}
\affiliation{Laboratoire J.L. Lagrange, UMR7293, Universit\'e de la C\^ote d'Azur, CNRS, Observatoire de la C{\^o}te d'Azur, B.P. 4229, 06304 Nice Cedex 4, France}
\author{Giorgio Krstulovic}
\affiliation{Laboratoire J.L. Lagrange, UMR7293, Universit\'e de la C\^ote d'Azur, CNRS, Observatoire de la C{\^o}te d'Azur, B.P. 4229, 06304 Nice Cedex 4, France}
\begin{abstract}
One of the main features of superfluids is the presence of topological defects with quantised circulation. These objects are known as quantum vortices and exhibit a hydrodynamic behaviour. Nowadays, particles are the main experimental tool used to visualise quantum vortices and to study their dynamics. We use a self-consistent model based on the three-dimensional Gross-Pitaevskii (GP) equation to explore theoretically and numerically the attractive interaction between particles and quantised vortices at very low temperature. Particles are described as localised potentials depleting the superfluid and following Newtonian dynamics. We are able to derive analytically a reduced central-force model that only depends on the classical degrees of freedom of the particle. Such model is found to be consistent with the GP simulations. We then generalised the model to include deformations of the vortex filament. The resulting long-range mutual interaction qualitatively reproduces the observed generation of a cusp on the vortex filament during the particle approach. Moreover, we show that particles can excite Kelvin waves on the vortex filament through a resonance mechanism even if they are still far from it.
\end{abstract}
%\maketitle
\flushbottom
\maketitle

\section{Introduction}

Quantum vortices have a long history in physics of superfluids and superconductors. Already in the 40's Onsager had suggested the existence of quantised flows. This idea  was further developed by Feynman by introducing the concept of quantum vortices \cite{donnelly1991quantized}. What makes these vortices fascinating is that they appear as topological defects of the order parameter describing the system. As a consequence their charge or circulation is quantised, making them topological protected objects.
 Their core size varies form a few Angstroms in superfluid $^4$He to micrometers in Bose-Einstein condensates (BECs).
In systems such as $^4$He, $^3$He and atomic BECs, quantum vortices behave as hydrodynamic vortices, reconnecting and rearranging their topology, forming in this way complex vortex tangles. Such out-of-equilibrium state is today known as quantum turbulence \cite{BarenghiIntroQunatumTurbu}. In rotating BECs, quantised vortices naturally appear and they have been studied since the early 2000s \cite{VortexStirredBEC,VortexLatticeOBS}
In superfluid helium, ions and impurities have been extensively used since long time to investigate the properties of quantum vortices \cite{donnelly1991quantized}. However, an important experimental breakthrough occurred in 2006 \cite{bewley2006superfluid}, when quantum vortices were directly visualised by using micrometer-sized hydrogen particles. These impurities are trapped inside the vortex core and they can be directly visualised by using standard particle-tracking techniques, that are commonly exploited in classical hydrodynamic turbulence. Thanks to this method, quantum vortex reconnections \cite{BewleyReconnectionExp} and Kelvin waves propagating along the vortex filaments \cite{FondaKWExp} have been observed. In addition, the employment of particles has been helpful to enlighten similarities and differences between classical hydrodynamic and quantum turbulence \cite{PaolettiVelStat2008,LaMantiaPRBVelStat2014}. For superfluid helium, the typical size of hydrogen particles is several orders of magnitude larger than the vortex core, whereas recent experiments have used He$^*_2$ excimers that are slightly larger than the vortex core \cite{ZmeevExcimers}. Therefore, understanding the interaction between particles and vortices has become crucial for current experiments.

In general, utilising particles to unveil the properties of a fluid is a common technique in classical hydrodynamics. For instance, air bubbles are used to visualise classical vortices in water since the pioneering work of Couder et al. in 1991 \cite{DouadyCouderBrachetVorticesPRL} and tracers (very small and neutrally-buoyant particles) are followed by using ultra-fast-cameras to determine the statistics of turbulent flows \cite{toschi2009lagrangian}. When particles are not tracers, they manifest inertia with respect to the fluid flow, deviating from its stream lines. Although complex, their dynamics is well understood in classical fluids if their size is small enough \cite{MaxeyNonUniform,AutonNonUniform}. 

Superfluids differ in several aspects to classical fluids. Firstly, at very low temperature, an object moving at low velocity experiences no drag. Secondly, the quantum nature of vortices makes the vorticity field (the curl of the velocity) a Dirac-$\delta$ distribution supported on the vortex filaments. Finally, at finite temperature, they are modelled  by an immiscible mixture of two components: the actual superfluid and a {\it normal} fluid. The latter is described by the (viscous) Navier-Stokes equations. Such mixture of fluids is responsible for some quantum effects with no classical analogous such as the fountain effect and second sound \cite{donnelly1991quantized}. The dynamics of a particle moving in a finite temperature superfluid happens to be richer than in an ordinary fluid. Its equations of motion have been generalised to the case where the flow is prescribed by the two-fluid model \cite{ParticlesNewcastle,ReviewNewcastle}. This model provides a large-scale description of a finite temperature superfluid where vortices are described with a coarse-grained field, therefore the quantised nature of superfluid vortices is missing. A different model that does account for the quantised nature of superfluid vortices, was introduced by Schwartz and it is  known as the vortex filament method \cite{SchwartzVFM}. Also in this case, the dynamics of particles has been addressed both theoretically and numerically \cite{SchwartzParticle,FiniteTemperatureFirstNewcastle}. Eventually, in the limit of very low temperature, superfluids can be described by another important model, the Gross-Pitaevskii (GP) equation. This model derives from a mean field approximation of a quantum system and directly applies to weakly-interacting BECs, but it is also expected to qualitatively apply to other types of superfluids. The GP equation governs the dynamics of the macroscopic wave function of the system, hence quantum vortices are naturally included. In the GP framework, impurities and particles are often described in terms of classical fields \cite{PitaevskiiTheory, CaptureBerloff,RicaRoberts,VilloisBubble}. In particular, it was shown by Roberts and Rica \cite{RicaRoberts} that, depending on the coupling constants, the impurity field separates from the condensate and the two fields become immiscible. In this regime, an impurity can be seen as a hard-core particle described with classical (Newtonian) degrees of freedom \cite{ActiveWiniecki, ShucklaSticking, ShuklaParticlesPRA2017}. Such approach is numerically much cheaper than the classical field description, and thus allow for simulations of a large number of particles \cite{giuriato2018clustering}. It also suitable for developing analytical predictions. 

In this Report we study numerically and analytically the interaction of quantum vortices and particles by using the Gross-Pitaevskii model coupled with a particle having classical degrees of freedom. We take advantage of the Hamiltonian structure of the system to derive a simplified model for the particle motion that it is then directly confronted with numerical simulations of the full GP model. In particular, we study the trapping of particles by a straight vortex, where an explicit analogy of a Newtonian central force problem can be established. The model is then generalised to describe the deformation of the vortex filament. The consequences of the long-range interaction between the particle and the filament are analytically studied and a prediction for the generation of Kelvin wave is obtained.

\section{Model for particles in a superfluid}
We consider a superfluid at very low temperature with one spherical particle of radius $\Rp$ and mass  $\Mp$ immersed in it. The superfluid is described by a complex field $\psi(\mathbf{x},t)$ and the particle classical degrees of freedom are its position $\mathbf{q}=(q_x,q_y,q_z)$ and momentum $\mathbf{p}=\Mp\dot{\mathbf{q}}=(p_x,p_y,p_z)$. The dynamics of the system is governed by the following Hamiltonian:
%\begin{widetext}
\begin{equation}
H=\frac{{\mathbf{p}}^2}{2 \Mp}+\int\left( \frac{\hbar^2}{2m} |\grad \psi |^2 +\frac{g}{2}|\psi|^4-\mu|\psi|^2 + \Vp(| \x -{\bf q} |)|\psi|^2 \right) \mathrm{d} \x ,
\label{Eq:HGP}
\end{equation}
%\end{widetext}
where $m$ is the mass of the fundamental bosons constituting the superfluid, $\mu$ is the chemical potential and the coupling constant $g=4 \pi  a_\mathrm{s} \hbar^2 /m$ depends on the $s$-wave scattering length $a_{\rm s}$. The potential $\Vp(|\mathbf{x}-\mathbf{q}|)\gg\mu>0$ is localised around $\mathbf{q}$ and it determines the shape of the particle. Its presence induces a full depletion of the superfluid around the position $\mathbf{q}$ up to a distance $\Rp$. The equations of motion for the field and the particle position are directly obtained by varying \eqref{Eq:HGP} and read
\begin{eqnarray}
i\hbar\dertt{\psi} = - \frac{\hbar^2}{2m}\nabla^2 \psi + \left( g|\psi|^2-\mu\right)\psi+\Vp(| \x -{\bf q} |)\psi \label{Eq:GPEParticles},&&
\Mp\ddot{\bf q} = - \int  \Vp(| \x -{\bf q}|) \nabla|\psi|^2\, \mathrm{d} \x. 
\end{eqnarray}
%\begin{gather}
%i\hbar\dertt{\psi} = - \frac{\hbar^2}{2m}\nabla^2 \psi + \left( g|\psi|^2-\mu\right)\psi+\Vp(| \x -{\bf q} |)\psi \label{Eq:GPE}\\
%\Mp\ddot{\bf q} = - \int  \Vp(| \x -{\bf q}|) \nabla|\psi|^2\, \mathrm{d} \x. \label{Eq:Particles}
%\end{gather}
%
The Hamiltonian (\ref{Eq:HGP}), the total superfluid mass $M=m\int|\psi|^2\,\mathrm{d}\mathbf{x}$ and the total momentum $\mathbf{P}=\frac{i\hbar}{2}\int(\psi\nabla\psi^*-\psi^*\nabla\psi)\,\mathrm{d}\mathbf{x}+\mathbf{p}$ are conserved quantities. The connection of Eq. \eqref{Eq:GPEParticles} with hydrodynamics is made through the Madelung transformation $\psi(\x)=\sqrt{{\rho(\x)}/{m}}\,e^{i\frac{m}{\hbar}\phi(\x)}$ that maps the GP model into the continuity and Bernoulli equations of a fluid of density $\rho$ and velocity $\mathbf{v}_\mathrm{s}=\nabla\phi$. 

In absence of the particle, the GP equation has a simple steady solution corresponding to a constant flat condensate $\psi_\infty=\sqrt{{\rho_\infty}/{m}}=\sqrt{\mu/g}$.  If \eqref{Eq:GPEParticles} is linearised about $\psi_\infty$, large wavelength waves propagate with the phonon (sound) velocity $c=\sqrt{g\rho_\infty/m^2}$ and dispersive effects take place at length scales smaller than the healing length $\xi=\sqrt{\hbar^2/2g\rho_\infty }$. 

Another important steady solution corresponds to a straight quantum vortex
\begin{equation}
\psi_{\rm v}(x,y,z)=\sqrt{\rho_{\rm v}(x,y)/m}\,e^{i \frac{m}{\hbar}\phi_{\rm v}(x,y)}. \label{EQ:quantumVortex}
\end{equation}
The vortex density $\rho_\mathrm{v}$ vanishes at $(0,0,z)$ and the phase is given by $\phi_{\rm v}=\frac{\hbar}{m}\kappa\varphi$, with $\varphi$ the angle in the $(x,y)$ plane and $\kappa$ a non-zero integer.
The corresponding velocity field $\mathbf{v}_{\rm v}$ satisfies
\begin{eqnarray}
\mathbf{v}_{\rm v}=\frac{\kappa\hbar}{m}\frac{\hat{\varphi}}{|\mathbf{x}_\perp|}
&{\rm and}&
\Gamma=\frac{1}{\kappa}\oint_{\mathcal{C}} \mathbf{v}_{\rm v}\cdot \mathrm{d}{\bf \mathbf{l}}=\frac{h}{m}= 2\pi \sqrt{2} c\xi,\label{Eq:VortexVelCirculation}
\end{eqnarray}
where $\hat{\varphi}$ is the azimuthal versor and $\mathbf{x_\perp}=(x,y,0)$. Similarly, in the following we will denote $\q_\perp=(q_x,q_y,0)$ and $q_\perp=|\mathbf{q}_\perp|$. The close path $\mathcal{C}$ surrounds the vortex, whose circulation is thus given by $\kappa\Gamma$. 
We will consider $\kappa=\pm 1$ because it is the only stable solution. Note that the vortex core size is given by the healing length $\xi$, $\rho_{\rm v}$ and $\mathbf{v}_\mathrm{v}$ are radial functions and $\rho_{\rm v}\to \rho_\infty$ away from the vortex \cite{pismen1999vortices}.

When a particle is present, the ground state (without vortices) corresponds to a flat condensate with a strong density depletion at places where $\Vp(|\mathbf{x}-\mathbf{q}|)>\mu$. A good approximation when $\Rp\gg\xi$ is given by the Thomas-Fermi ground state that is obtained neglecting the kinetic term. It reads
\begin{equation}
\rho_{\rm p}(\x; \q)=\rho_{\rm p}(|\x-\q|)=\rho_\infty (1-\frac{\Vp(|\mathbf{x}-\mathbf{q}|)}{\mu})\theta[1-\frac{\Vp(|\mathbf{x}-\mathbf{q}|)}{\mu}], \label{Eq:RhoPart}
\end{equation}
with $\theta$ the Heaviside function. The size of the particles is thus roughly determined by the relation $\Vp(\Rp)\approx \mu$. The results presented in this work are independent of the functional shape of $\Vp$, provided that it is isotropic.

In numerics, we express the particle mass as $\Mp=\mathcal{M}M_0$, where $M_0$ is the mass of the displaced superfluid. Therefore, neutral-mass particles have $\mathcal{M}=1$, heavy particles have $\mathcal{M}>1$ and light particles have $\mathcal{M}<1$. Lengths are expressed in units of $\xi$, times in units of $\tau=\xi/c$, velocities in units of $c$ and energies are normalised by $M_0c^2$. Details on the numerical implementation and the particular choice of $\Vp$ are given in Appendix \ref{App:Num}.

\section{Interaction between particles and quantum vortices}

\begin{figure}[ht]
\centering
\includegraphics[width=1.\linewidth]{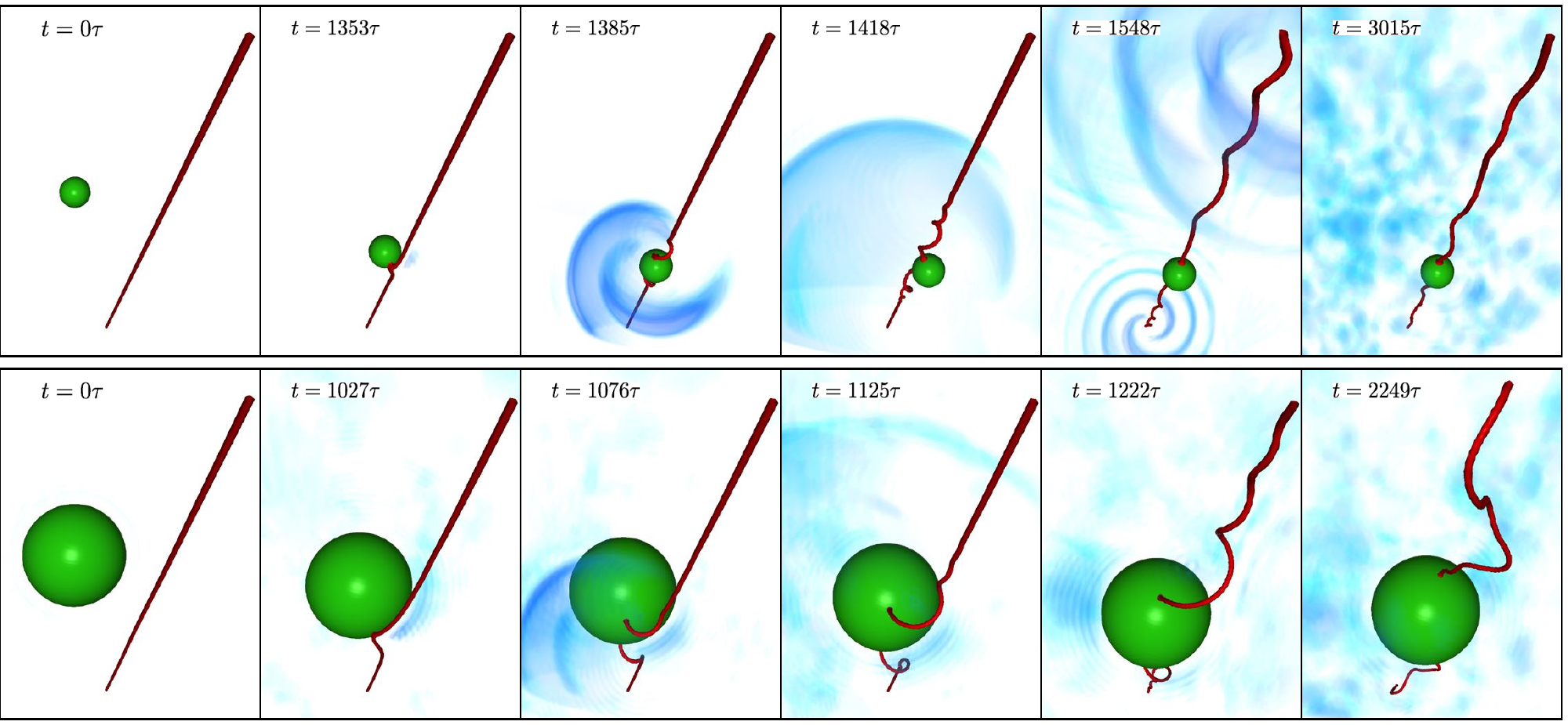}
\caption{Snapshots of the superfluid density and a neutral-mass particle during the trapping (times varies from left to to right). Vortices are displayed in red, particles in green and sound waves are rendered in blue. {\it Top}: small particle ($\Rp=7.6\xi$). {\it Bottom}: large particle ($\Rp=23.5\xi$). Images were produced with VAPOR rendering software.}
\label{Fig:3Dsketch}
\end{figure}

We begin by presenting some numerical experiments where a particle is attracted and captured by a vortex. We integrate the model \eqref{Eq:GPEParticles} in a 3D periodic domain of size $L=256\xi$ with an initial condition consisting of one particle at rest and one straight vortex initially separated by a distance $q_\perp=q_0\gg\xi$. The domain contains image vortices in order to preserve periodicity that are not displayed in figures. Their effect on the particle has been checked to be negligible. 
Snapshots of the superfluid density field with the particle at different times are displayed in Fig.\ref{Fig:3Dsketch}. The top row refers to a relatively small particle ($\Rp=7.6\xi$), while the bottom row to a large one ($\Rp=23.5\xi$). Both particles have a neutral relative mass $\mathcal{M}=1$. 
Note that hydrogen particles used for visualization of quantum vortices in superfluid helium have a relative mass {$\mathcal{M}\sim0.7$} and a typical size of $\Rp\sim 10^3\xi$. Simulating such particle size is not achievable numerically, however a clear difference is already observed for our large particle. 
In both cases, the particle is attracted by the vortex. Before the merging, while the particle is moving closer to the vortex, a deformation of the vortex line is observed. Such deformation is a cusp regularised at the scale of the healing length by the dispersion of the GP equation. Initially, the cusp develops perpendicularly to the particle velocity. Later, it curves towards the particle, until the contact point the vortex separates into two branches. The two contact points then slide on the particle surface towards opposite directions. The oscillation of the trapped particle excites helicoidal waves on the filament. Such waves, that propagate along the vortex line, are known as Kelvin waves. We note that the vortex deformation is less marked for smaller particles and the amplitude of Kelvin waves increases with the particle size. A similar behaviour has been already observed in the hydrodynamical model adopted in references \cite{ParticlesNewcastle,NumericsNewcastle,CloseParticleNewcastle}, as well as in the classical field impurity model studied by Berloff and Roberts \cite{CaptureBerloff}. The trapping process was then interpreted as a reconnection of the straight vortex with its images inside the particle, whose presence is necessary to set the boundary conditions for the flow around the particle.

The model \eqref{Eq:GPEParticles} also allow us to observe the sound emitted by the particle-vortex pair during the trapping process. In a first stage, a big pulse is emitted at the moment of the trapping due to the strong acceleration experienced by the particle. In a later time, for the smaller particle a clear quadrupolar radiative pulse is observed ($t=1548\tau$, top row of Fig. \ref{Fig:3Dsketch}). Remarkably, this kind of pattern is expected in superfluids when some symmetry cancels the first order of the multipolar radiative expansion. For instance, this is the case in 2D counter-rotating vortices \cite{Barenghi2005Radiation,KrstulovicRadiation}. Here, the symmetry could be related to the two antisymmetric traveling waves emerging from the particle and meeting at the boundary of the periodic domain. This issue will be investigated further in a future work. Finally, the particle remains trapped inside the vortex and coexists with a bath of sound waves. For the big particle, all the phenomena are amplified. Movies of the numerical simulations can be found as a Supplementary Information.

From these simulations it is manifest that the trapping of a particle by a quantum vortex is accompanied by a myriad of complex physical phenomena. In the next sections, we take advantage of the simplicity of the model to derive effective equations for the particle and the vortex Filament dynamics.

\subsection{Reduced theoretical model for the particle-vortex interaction}

In the following we set the origin of the reference frame at the intersection between the unperturbed vortex line and its orthogonal plane where the particle lies. At $t=0$, the vortex line coincides then with the $z$ axis. 
 To derive a simplified theory, we consider the following ansatz for the superfluid field: 
\begin{equation}
\psi(\mathbf{x};\q,\dot{\q})=\sqrt{\frac{\rho_\infty}{m}}\sqrt{\tilde{\rho}_\mathrm{v}(\mathbf{x})}\sqrt{\tilde{\rho}_{\rm p}( |\x-\q|   )}e^{i\frac{m}{\hbar}\phi(\mathbf{x};\q,\dot{\q})},
\label{Eq:ansatz}
\end{equation} 
where $\tilde{\rho}_\mathrm{v}=\sqrt{\rho_\mathrm{v}/\rho_\infty}$ and $\tilde{\rho}_\mathrm{p}=\sqrt{\rho_\mathrm{p}/\rho_\infty}$ are the normalised ground states of an isolated vortex and an isolated particle given in \eqref{EQ:quantumVortex} and \eqref{Eq:RhoPart} respectively. 
At first approximation, we neglect the deformation of the vortex. This last assumption is valid at the stages where the particle is attracted by the vortex, but still far from it. We will consider the vortex deformation in the last section.
The ansatz \eqref{Eq:ansatz} also neglects small density variations due to sound emission and might not be valid at the exact moment of the trapping, but it gives a good description elsewhere. The phase $\phi$ leads to the superfluid velocity field $\mathbf{v}_\mathrm{s}=\nabla\phi$ and it is determined by imposing the boundary conditions around the particle and at infinity:
\begin{equation}
\mathbf{\dot{q}}\cdot\mathbf{n}=\mathbf{v}_\mathrm{s}\cdot\mathbf{n}\qquad\forall\,\mathbf{x}\,\,\mathrm{s.t.}\,\,\,{|\mathbf{x-\q}|=\Rp} \qquad {\rm and} \qquad \mathbf{v}_\mathrm{s}\underset{|\mathbf{x}-\q|\to\infty}{\longrightarrow}\mathbf{v}_{\rm v}.
\label{Eq:boundary}
\end{equation}
where $\mathbf{n}=(\mathbf{x-\q})/|\mathbf{x}-\q|$ and $\mathbf{v}_{\rm v}$ the vortex velocity field \eqref{Eq:VortexVelCirculation}. 
Since $\mathbf{v}_{\rm v}$ describes a non-uniform irrotational flow, we have to take into account how the superfluid velocity field is modified when the particle accelerates in it. As it is done in classical fluid mechanics \cite{AutonNonUniform,MaxeyNonUniform,ParticlesNewcastle}, we include in the superfluid velocity the corrections to the pure vortex flow $\mathbf{v}_\mathrm{v}$ that are generated by the moving particle. We set $\mathbf{v}_\mathrm{s}=\mathbf{v}_\mathrm{v}+\mathbf{v}_\mathrm{p}+\mathbf{v}_\mathrm{BC}$, or in terms of the phase $\phi=\phi_{\rm v}+\phi_\mathrm{p}+\mathrm{\phi_\mathrm{BC}}.$

The potential $\phi_\mathrm{p}$ describes the flow of a sphere of radius $\Rp$ moving in a uniform flow given by the relative velocity $\mathbf{\dot{q}}-\mathbf{v}_{\rm v}(\q).$ It reads \cite{BatchelorClassic}
\begin{equation}
\phi_\mathrm{p}(\x;\q,\dot{\q})=-\frac{\Rp^3}{2|\mathbf{x-\q}|^3}(\mathbf{x-\q})\cdot(\mathbf{\dot{q}}-\mathbf{v}_{\rm v}(\mathbf{q})).
\label{Eq:phase_p}
\end{equation}
The potential $\phi_\mathrm{BC}$ is in principle determined by the condition at the particle boundary $\nabla\phi_\mathrm{BC}\cdot\mathbf{n}=\left[\mathbf{v}_{\rm v}(\mathbf{q}+\Rp\mathbf{n})-\mathbf{v}_{\rm v}(\mathbf{q})\right]\cdot\mathbf{n}$. In practice, it is obtained by a Taylor expansion of the vortex velocity flow around the particle and hence $\phi_\mathrm{BC}$ is expressed in terms of its gradients \cite{AutonNonUniform,MaxeyNonUniform,ParticlesNewcastle}. This flow gives a contribution of order
\begin{equation}
\epsilon=\Rp\frac{|\nabla\mathbf{v}_{\rm v}(\q)|}{|\mathbf{v}_{\rm v}(\mathbf{q})|}=\frac{\Rp}{q_\perp}\ll 1,
\label{Eq:epsilon}
\end{equation}
where we have used $|\mathbf{v}_{\rm v}(\mathbf{q})|\sim1/q_\perp$. We include in our calculations $\phi_\mathrm{BC}$ up to $\mathcal{O}(\epsilon^2)$.

In order to express the Hamiltonian \eqref{Eq:HGP} only in terms of $\q$ and $\dot \q$ we split it as 
$H=K+H^\mathrm{GP}_\mathrm{hydro}+H^\mathrm{GP}_\mathrm{int} +H^\mathrm{GP}_\mathrm{p}+H^\mathrm{GP}_\mathrm{qnt}$, where
\begin{eqnarray}
&K=\frac{1}{2}\Mp\dot{\mathbf{q}}^2,\qquad
H^\mathrm{GP}_\mathrm{hydro}=\frac{1}{2}\int\rho\mathbf{v}_\mathrm{s}^2\,\mathrm{d}\mathbf{x},\qquad
H^\mathrm{GP}_\mathrm{int}=\frac{g}{2m^2}\int\rho^2\,\mathrm{d}\mathbf{x}\nonumber,\\
&H^\mathrm{GP}_\mathrm{p}=\frac{1}{m}\int\left(V_\mathrm{p}-\mu\right)\rho\,\mathrm{d}\mathbf{x},\qquad
H^\mathrm{GP}_\mathrm{qnt}=\frac{\Gamma^2}{8\pi^2}\int\left(\nabla\sqrt\rho\right)^2\,\mathrm{d}\mathbf{x}.
\label{Eq:H_terms}
\end{eqnarray}
We use the ansatz \eqref{Eq:ansatz} to explicitly perform the space integrals. 
From \eqref{Eq:RhoPart}, we observe that for a strong localised potential $V_{\rm p}\gg\mu$ the field $1-\tilde{\rho}_\mathrm{p}(\x)$ is supported on a ball of center $\q$ and radius $\Rp$, up to a layer of size $\xi$. We use this fact to reduce the domain of integration. Inside this ball and if $\epsilon\ll 1$, we can assume that $\tilde{\rho}_\mathrm{v}(\mathbf{x})\approx\tilde{\rho}_\mathrm{v}(q_\perp)$ and $\mathbf{v}_\mathrm{v}(\mathbf{x})\approx\mathbf{v}_\mathrm{v}(q_\perp)$. All the integrations can be then carried out. Details on these computations are given in Appendix \ref{App:Model}.
The Hamiltonian components (\ref{Eq:H_terms}) eventually read
\begin{eqnarray}
H^\mathrm{GP}_\mathrm{hydro}&\approx&
\bar{H}_\mathrm{hydro}^{GP}-\frac{\left(1+C\right)M_0\Gamma^2}{8\pi^2 q_\perp^2}\tilde{\rho}_\mathrm{v}(q_\perp)-\frac{\Gamma^2 \Rp^2}{20\pi^2 c^2 q_\perp^4}\tilde{\rho}_\mathrm{v}(q_\perp)+ E_\mathrm{add},
{\qquad\rm with\qquad}E_\mathrm{add}=\frac{1}{2}C\tilde{\rho}_\mathrm{v}(q_\perp)M_0\mathbf{\dot{q}}^2,\label{Eq:Hhydro}\\
H^\mathrm{GP}_\mathrm{int}&\approx&
\bar{H}^\mathrm{GP}_\mathrm{int}-\frac{1}{2}M_0c^2\tilde{\rho}_\mathrm{v}^2(q_\perp),\qquad 
H^\mathrm{GP}_\mathrm{p}\approx
\bar{H}^\mathrm{GP}_\mathrm{p}+M_0c^2\tilde{\rho}_\mathrm{v}(q_\perp),
\qquad H^\mathrm{GP}_\mathrm{qnt}\approx\bar{H}^\mathrm{GP}_\mathrm{qnt}
\label{Eq:HintHp}
\end{eqnarray}
where $C=1/2$ and overbars denote constants at the leading order. $E_\mathrm{add}$ is the classical added mass energy in three dimensions \cite{BatchelorClassic} modified by the density profile. 
Gathering all the terms, we obtain the reduced Hamiltonian (RH)
\begin{equation}
H_\mathrm{red}[\mathbf{q},\mathbf{p}]=
\bar{H}^\mathrm{GP}+\frac{\mathbf{p}^2}{2M_\mathrm{eff}}
+M_0c^2\left[-\frac{1}{2}\tilde{\rho}_\mathrm{v}^2(q_\perp) + \tilde{\rho}_\mathrm{v}(q_\perp) 
- \frac{\left(1+C\right)\Gamma^2}{8\pi^2 c^2 q_\perp^2}\tilde{\rho}_\mathrm{v}(q_\perp)
- \frac{\Gamma^2 \Rp^2}{20\pi^2 c^2 q_\perp^4}\tilde{\rho}_\mathrm{v}(q_\perp)
\right].
\label{Eq:HGPred}
\end{equation}
In (\ref{Eq:HGPred}) the added mass has been absorbed in the effective particle mass $M_\mathrm{eff}=\Mp+C\tilde{\rho}_\mathrm{v}M_0=(\mathcal{M}+C\tilde{\rho}_\mathrm{v}) M_0$ and the particle momentum has been redefined as $\mathbf{p}=M_\mathrm{eff}\mathbf{\dot{q}}$. 
Note that, as $\tilde{\rho}_\mathrm{v}$ only depends on $q_\perp$, the coordinate $q_z$ of the particle is cyclic and can be trivially integrated.
The dynamics thus simplifies to a motion in the plane perpendicular to the vortex. 
The reduced model \eqref{Eq:HGPred}, therefore describes a classical central force problem in two dimensions with a potential given by its last term.
Note that the same calculations can be performed in two dimensions, by redefining the phase (\ref{Eq:phase_p}) which leads to the constant $C=1$. 

The reduced Hamiltonian \eqref{Eq:HGPred} can be further simplified using the asymptotic behaviour of $\tilde{\rho}_\mathrm{v}$. At large distances,  $\tilde{\rho}_\mathrm{v}\sim1-\xi^2/q_\perp^2-2\xi^4/q_\perp^4+O(\xi^6/q_\perp^6)$, hence at the leading order $H^\mathrm{GP}_\mathrm{p}\approx - H^\mathrm{GP}_\mathrm{int}$ and the main contribution comes only from $H^\mathrm{GP}_\mathrm{hydro}$. Finally, at lowest order, we obtain the effective Hamiltonian (EH) for the particle dynamics
\begin{equation}
H_\mathrm{eff}[\q,\p]=\frac{\mathbf{p}^2}{2M_\mathrm{eff}}+U(q_\perp),\quad U(q_\perp)=-\frac{\left(1+C\right)M_0\Gamma^2}{8\pi^2 q_\perp^2}, \quad{\rm and}\,\quad
M_\mathrm{eff}=\Mp+CM_0=(\mathcal{M}+C) M_0.
\label{Eq:Heff}
\end{equation}
The equations of motion for the particle position are then:
\begin{equation}
\left(\mathcal{M}+C\right)\ddot{\mathbf{q}}_\perp=-\frac{\left(1+C\right)\Gamma^2}{4\pi^2q_\perp^4}\mathbf{q_\perp},\qquad\ddot{q}_z=0,
\label{Eq:motion}
\end{equation}
Note that the added mass effect is suppressed for neutral-mass particles having $\mathcal{M}=1$ and the particle size explicitly appears only at high order terms.
The attractive force scaling as $q_\perp^{-3}$ was first proposed by Donnelly \cite{Donelly65PresureGrad} as the result of a pressure gradient. Equation \eqref{Eq:motion} has been also studied for neutral-mass particles in the framework of pure hydrodynamical models \cite{ParticlesNewcastle,ReviewNewcastle}.

Finally, note that if we replace in \eqref{Eq:HGPred} the density by its leading order $\tilde{\rho}_\mathrm{v}=1$, then the associated equations of motion are invariant under the following scaling transformation:
\begin{equation}
\mathbf{q}\rightarrow \lambda\mathbf{q},\quad a_\mathrm{p}\rightarrow \lambda a_\mathrm{p},\quad t\rightarrow \lambda^2 t\qquad \forall \lambda\in\mathbb{R}^+
\label{Eq:scaling}
\end{equation}
Such invariance will be also preserved in terms coming from higher orders in $\epsilon$.

\subsection{Numerical measurements and comparison with theory}

\begin{figure}[ht]
\centering
\includegraphics[width=1.\linewidth]{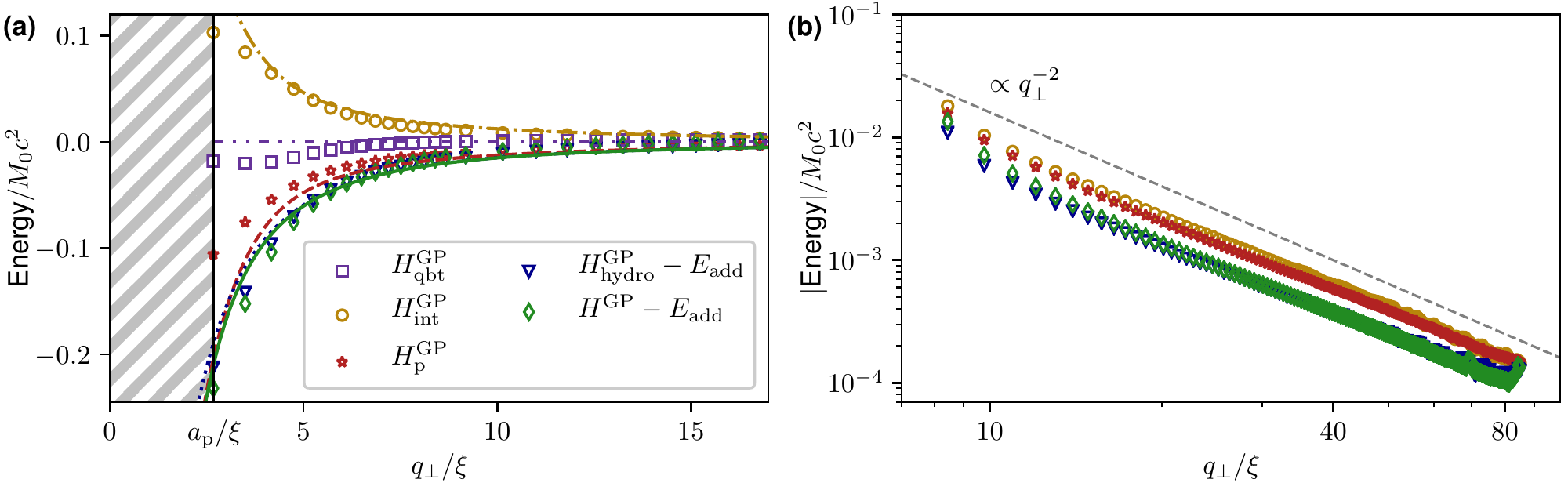}
\caption{\textbf{(a)} Different energies as a function of the vortex-particle distance $q_\perp/\xi$ during the approach of a particle with size $\Rp=2.7\xi$. The initial separation is $q_\perp=45.3\xi$ and the particle has zero velocity. Markers are numerical data and lines theoretical curves of corresponding colours. \textbf{(b)} Same energies as in \textbf{(a)} in ($\log -\log$ scale). Initial $q_\perp=85.1\xi$ and initial velocity $\dot{q}_\perp=-0.04c$.}
\label{Fig:TrappingEnergies}
\end{figure}

We compare now our reduced model to the numerical experiment presented in Fig. \ref{Fig:3Dsketch}. We first consider a small neutral-mass particle of size $\Rp=2.7\xi$. For this particle, the condition (\ref{Eq:epsilon}) is valid for a wide range of separations and the deformation of the vortex during the particle approach is negligible. 
We measured the variation of the different components of the Hamiltonian (\ref{Eq:H_terms}) as a function of distance between the particle and the vortex. Fig. \ref{Fig:TrappingEnergies}a displays such energies (markers) compared to the respective theoretical predictions (lines).
The striped region identifies the particle radius $\Rp/\xi$ where the particle and the vortex overlap. Since the added mass energy $E_\mathrm{add}$  (\ref{Eq:Hhydro}) only modifies the particle inertia but has no effect in determining the force in the r.h.s. of (\ref{Eq:motion}), we subtract it from the hydrodynamic component $H^\mathrm{GP}_\mathrm{hydro}$ and the total GP energy $H^\mathrm{GP}=H-K$. 
We have used the Pad\'e approximation given in Appendix \ref{App:Num} as an analytical expression for the vortex density profile $\tilde{\rho}_\mathrm{v}$, so that both asymptotics (large and short vortex-particle separations) are reproduced. Even if our model is not supposed to be quantitatively accurate for  $q_\perp\sim\Rp$, we can still observe a quite good agreement. Remarkably, the hypothesis that leads to neglect $H^\mathrm{GP}_\mathrm{q}$ is perfectly valid up to a distance about twice of the particle radius. Moreover, $H^\mathrm{GP}_\mathrm{p}\approx - H^\mathrm{GP}_\mathrm{int}$ and thus during the particle approach $H^\mathrm{GP}\approx H^\mathrm{GP}_\mathrm{hydro}$.
Fig. \ref{Fig:TrappingEnergies}b shows in a $\log -\log$ plot the absolute value of the measured energies for large distances. Clearly, all the energy contributions follow the predicted $q_\perp^{-2}$ scaling, as long as the vortex-particle separation is large. We have checked that the data in Fig. \ref{Fig:TrappingEnergies} are almost independent of the particle mass. Discrepancies between data and theory might be due to sound radiation or to sub-leading terms in the boundary conditions of the superfluid velocity. We conclude that the effective potential energy is relatively well described by $U(q_\perp)$ in Eq.\eqref{Eq:Heff} and Eq.(\ref{Eq:motion}) gives a good approximation for the motion of the particle.

\begin{figure}[ht]
\centering
\includegraphics[width=1.\linewidth]{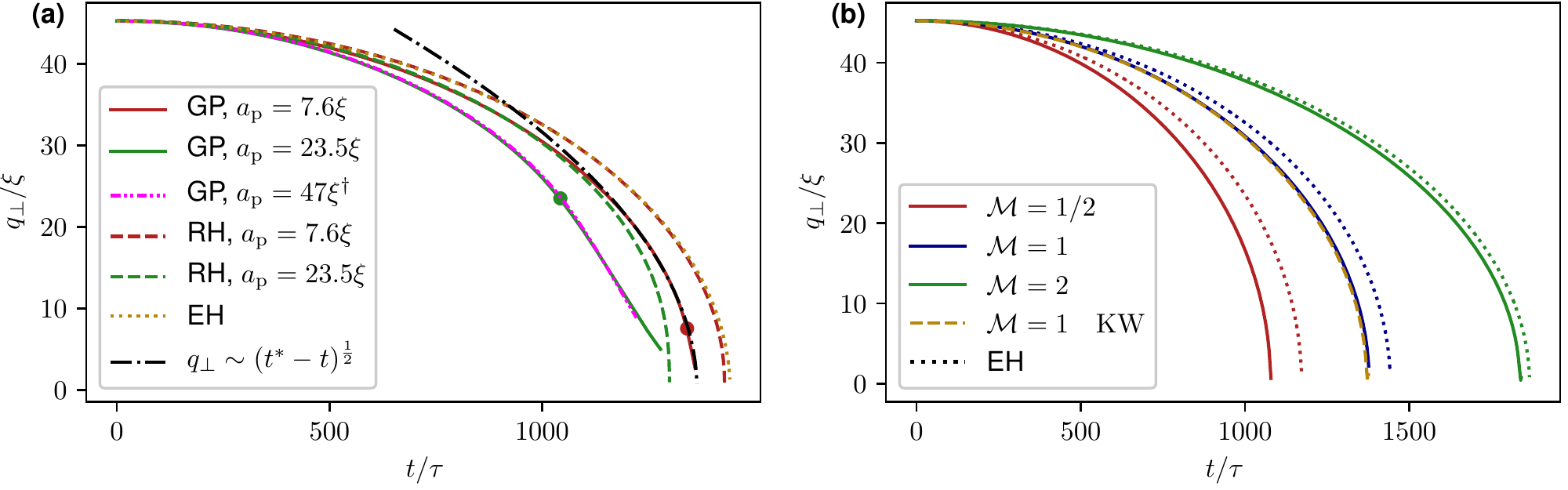}
\caption{\textbf{(a)} Measured vortex-particle separation as a function of time for neutral-mass ($\mathcal{M}=1$) particles of different sizes moving towards a straight vortex (solid lines). The initial condition is $q_\perp=45.3\xi$ and $\dot{\q}=0$. Round markers indicate the corresponding trapping times. $^\dagger$ The pink dash-dot-dotted line refers to a big particle initially at $q_\perp=90.6\xi$ rescaled using Eq.\eqref{Eq:scaling} with $\lambda=2$. The figure also displays the predictions RH (\ref{Eq:HGPred}) in dashed lines of the corresponding colours, the theoretical prediction (\ref{Eq:solution}) of the effective model EH (dotted golden line) and the scaling $q\sim(t^*-t)^\frac{1}{2}$, obtained by fitting the numerical data (dash-dotted black line). \textbf{(b)} The same as \textbf{(a)} but for particles with $\Rp=2.7\xi$ and different masses. Data from GP simulation are displayed in solid lines whereas the predictions (\ref{Eq:solution}) are in dotted lines. The dashed golden line refers to a GP simulation with a vortex containing Kelvin waves of {\it rms} amplitude $0.5\xi$.}
\label{Fig:TrappingMotion}
\end{figure}

Equation (\ref{Eq:motion}) can be straightforwardly integrated and the solution for the particle-vortex distance reads
\begin{equation}
q_\perp(t)=\sqrt{\frac{2E_\perp}{M_\mathrm{eff}}t^2+q_0^2+ q_0\dot{q}_0t},
\label{Eq:solution}
\end{equation}
where $q_0=q_\perp(t=0)$, $\dot{q}_0=\dot{q}_\perp(t=0)$ and $E_\perp=H_\mathrm{eff}[\mathbf{q}(t={0}),\mathbf{p}(t={0})]-p_z^2/2M_\mathrm{eff}$ is the conserved energy in the effective model. In the case of a neutral-mass particle with zero initial velocity Eq. (\ref{Eq:solution}) reduces to the one derived by Barenghi et al.\cite{CloseParticleNewcastle}. In Fig. \ref{Fig:TrappingMotion} the prediction (\ref{Eq:solution}) and the one obtained numerically from RH \eqref{Eq:HGPred} are compared with numerical data. 
Fig. \ref{Fig:TrappingMotion}a shows the particle-vortex distance for neutral-mass particles of different sizes initially located at a distance $q_0=45.3\xi$. The markers denote the capture times, after which particles keep moving inside the vortex. The assumption (\ref{Eq:epsilon}) is ideally satisfied for point-like particles but it reasonably applies as long as the particle radius is sufficiently small compared to its distance to the vortex. Indeed, for the particle $\Rp=7.6\xi$ the accordance with theory is good, while for the one having $\Rp=23.5\xi$ is just qualitative. For such particle, the full reduced Hamiltonian gives a better description. In addition, the motion curve of a particle of radius $\Rp=2\times23.5\xi$, initially located at $2q_0$ is in good agreement (pink dashed-dot-dotted curve) with the scaling relation (\ref{Eq:scaling}).
It is interesting to note that close to the capture time the particle-vortex separation scales as
\begin{equation}
q(t)\underset{t\rightarrow t^{*-}}{\longrightarrow} \left(  \frac{1+C}{\pi^2(\mathcal{M}+C)}\right)^{1/4}\sqrt{\Gamma(t^*-t)},\qquad\mathrm{with}\quad t^*=\frac{ q_0^2}{\Gamma}\sqrt{\frac{4\pi^2(\mathcal{M}+C)}{1+C}},
\label{Eq:closet}
\end{equation}
where we have set $\dot{\q}(t=0)=0$ for sake of simplicity. In Fig. \ref{Fig:TrappingMotion}a such scaling is also apparent up to a separation of $q\sim 30\xi$ for the particle with $\Rp=7.6\xi$ (dotted-dashed line). The capture time $t^*$ predicted by the effective model is compatible with the one observed in the GP simulation with a relative error of $5\%$. The scaling (\ref{Eq:closet}), that is also observed in vortex reconnections\cite{ReconnectionGiorgio}, suggests the idea that the trapping process could be seen as reconnection of the vortex with its images inside the particle.
Finally, in Fig. \ref{Fig:TrappingMotion}b the vortex-particle separation has been measured for a small particle ($\Rp=2.7\xi$) with the same initial condition but with different masses. Remarkably, the heavier the particle, the better the agreement with theory. This could due to the fact that light particles are more sensitive to sound waves and compressible effects not taken into account in the theory. For completeness, we also show the case of a vortex filament perturbed with small-amplitude Kelvin waves (dashed golden line). As expected, the effect of Kelvin waves is sub-leading and no difference is appreciable with respect to the unperturbed case.

\begin{figure}[ht]
%\centering
\includegraphics[width=1.\linewidth]{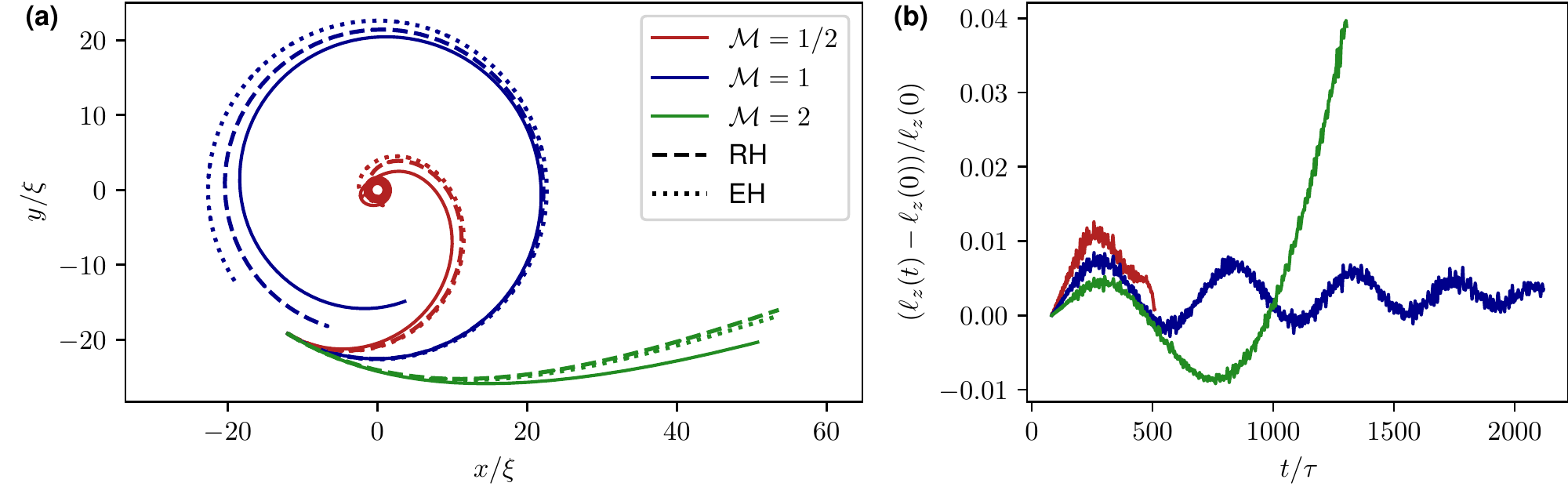}
\caption{\textbf{(a)} Trajectories of small particles ($a_\mathrm{p}=2.7\xi$) with $\mathbf{\dot{q}}_\perp(t=0)=\mathbf{v}_\mathrm{v}(q_\perp)$ and $q_\perp(t=0)=22.6\xi$. GP data are displayed in solid lines, predictions of EH (\ref{Eq:Heff}) in dotted lines and the predictions of RH (\ref{Eq:HGPred}) in dashed lines of the corresponding colours. \textbf{(b)} Relative variation of angular momentum as a function of time for the same simulations of \textbf{(a)}.}
\label{Fig:Orbits}
\end{figure}

It is well known in classical hydrodynamics that light particles go into vortices whereas the heavy ones escape from them \cite{DouadyCouderBrachetVorticesPRL}. 
The same situation takes place for a particle in a superfluid, even if there is no Stokes drag at zero temperature. Indeed, as a central force problem, the effective Hamiltonian (\ref{Eq:Heff}) conserves the angular momentum $\ell_z=M_\mathrm{eff} (\q_\perp \times \dot{\q}_\perp)\cdot \hat{z}$. This conserved quantity leads to the emergence of a repulsive potential $\ell_z^2/2M_\mathrm{eff}\,q_\perp^2$ in the effective Hamiltonian for $q_\perp$. Therefore there exists a critical angular momentum $\ell_\mathrm{crit}=\sqrt{(1+C)(\mathcal{M}+C)}M_0\Gamma/2\pi$ such that for $\ell_z<\ell_\mathrm{crit}$ particles collapse into the vortex and escape from it for $\ell_z>\ell_\mathrm{crit}$. 
Now, if the particle is initially at rest in the reference frame moving with vortex flow, i.e. $\dot{\q}_\perp={\bf v}_{\rm v}(q_\perp)$, the condition on the critical value of $\ell_z$ is expressed in terms of the mass, as $\mathcal{M}<1$ for trapping and $\mathcal{M}>1$ for escaping. At $\mathcal{M}=1$ the model (\ref{Eq:Heff}) predicts a closed circular orbit, i.e. a particle tracing the flow. However, this orbit is unstable and modified by high order terms (see \ref{App:Model}) that lead to a collapse also in this case. The three situations $\mathcal{M}<1$, $\mathcal{M}=1$ and $\mathcal{M}>1$ are manifest in Fig. \ref{Fig:Orbits}a, where we display the trajectories of a small particle ($a_\mathrm{p}=2.7\xi$) with initial velocity $\dot{\q}_\perp={\bf v}_{\rm v}(q_\perp)$ but different masses. For the $\mathcal{M}=1$ case the prediction given by (\ref{Eq:HGPred}) works better than the leading order solution. This is consistent with the fact that the terms proportional to $q_\perp^{-2}$ cancel for $\ell_z=\ell_\mathrm{crit}$, so that the next-to-leading order becomes predominant.
Fig. \ref{Fig:Orbits}b shows that the angular momentum is conserved up to $4\%$. Note that the escaping particle feels the attraction of image vortices in the periodic box that break down the conservation of $\ell_z$.

\section{Generation of cusps and Kelvin waves on the vortex filament}

\begin{figure}[ht]
\centering
\includegraphics[width=1.\linewidth]{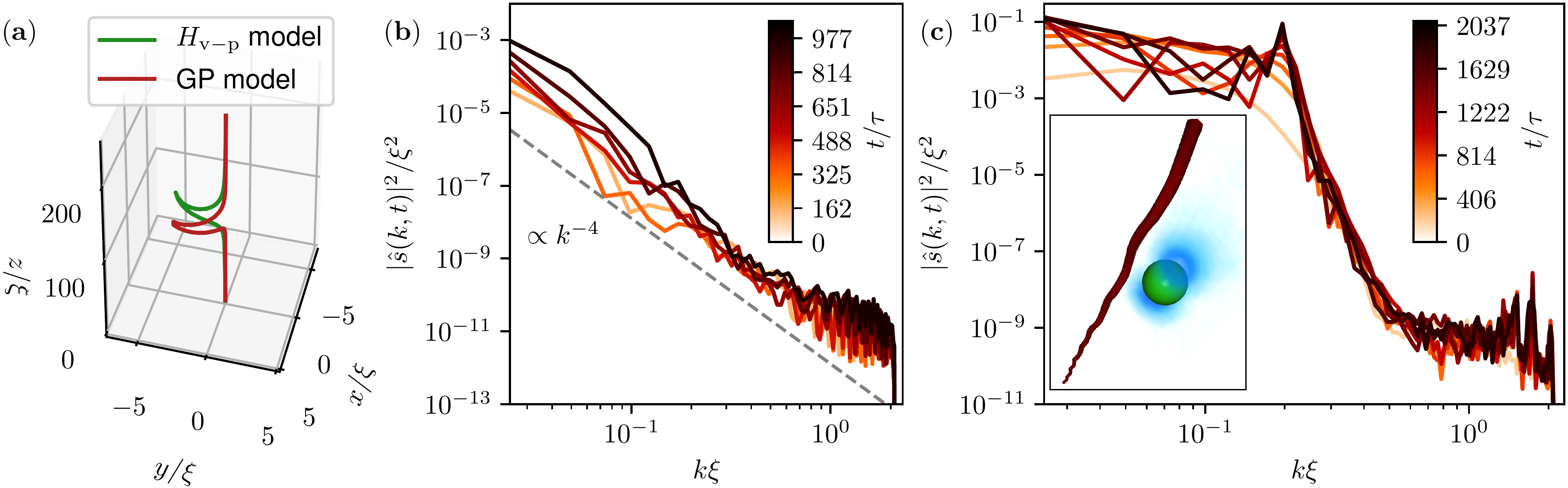}
\caption{\textbf{(a)} Cusp generated during the trapping of a neutral-mass particle of size $a_\mathrm{p}=7.6\xi$ at the capture time $t^*$. The initial particle condition is $q_\perp=45.3\xi$ and $\dot{\q}=0$. Red solid line is the vortex line tracked during GP simulation, whereas green solid line is $\mathbf{s}(z,t^*)$ computed with the dynamics (\ref{Eq:VortexParitlceMotion1}-\ref{Eq:VortexParitlceMotion2}). \textbf{(b)} Spectrum of vortex displacement measured from GP simulation at different times during the trapping of a particle. The parameters are the same as in \textbf{(a)}. \textbf{(c)} Spectrum of vortex displacement measured from GP simulation during the motion of a particle. The parameters for the particle are $a_\mathrm{p}=7.6\xi$, $q_\perp(t=0)=22.6\xi$, $\dot{\q}_\perp(t=0)=\mathbf{v}_\mathrm{v}(q_\perp)$ and $\dot{q}_z=0.27c$ (see Appendix \ref{App:Num}). The inset shows the superfluid density and the particle where Kelvin waves are clearly present on the vortex filament.}
\label{Fig:Cusp}
\end{figure}

We now address the effect of the particle on the vortex filament. As the vortex remains almost straight, it can be parametrised as ${\bf R}(z)=({\bf s}(z),z)$, where ${\bf s}(z)$ is a bi-dimensional vector. The ansatz \eqref{Eq:ansatz} can be generalised by replacing $\x_\perp$ in $\rho_{\rm v}$ and $\phi_{\rm v}$ by $\x_\perp - {\bf R}(z)$. Assuming $|{\bf s}(z)|\ll q_\perp$ and small deflections $|\partial_z {\bf s}|\ll 1$, all the calculations made in the previous section to reduce the Hamiltonian can be performed in the same way if we keep only contributions at the first order in ${\bf s}(z)$. The vortex deformation appears in the term $\bar{H}^\mathrm{GP}$ in \eqref{Eq:HGPred} and simply corresponds to the Local Induced Approximation (LIA) Hamiltonian\cite{pismen1999vortices} (see Appendix \ref{App:Vortex}). The effective vortex-particle Hamiltonian \eqref{Eq:Heff} becomes
\begin{equation}
H_\mathrm{v-p}[\q,\p,{\bf s}]=\frac{\mathbf{p}_{\mathrm{eff}}^2}{2M_\mathrm{eff}}+\frac{\Gamma^2\rho_\infty}{8\pi}\int_0^L\left[ - \mathbf{s}\cdot \Lambda\frac{\partial^2\mathbf{s}}{\partial z^2}-\frac{(1+C)M_0}{\rho_\infty\pi|\mathbf{q_\perp}-\mathbf{s}(z)|^2}\delta(z-q_z)\right]\,\mathrm{d}z,\label{Eq:HeffV}
\end{equation}
where from now on $\q_\perp=(q_x,q_y)$. In principle $\Lambda$ is a non-local operator yielding the correct Kelvin wave dispersion relation\cite{pismen1999vortices}. For the moment, we treat it as a constant. Up to a logarithmic correction, this is equivalent to consider the limit of large-scale vortex deformations. Although rough, such approximation provides a qualitatively good description of the vortex dynamics. The equations of motion coupling the vortex filament and the particle are thus found to be:
\begin{eqnarray}
\left(\mathcal{M}+C\right)\ddot{\mathbf{q}}_\perp&=&-\frac{\left(1+C\right)\Gamma^2}{4\pi^2|\mathbf{q_\perp}-\mathbf{s}(q_z)|^4}(\mathbf{q_\perp}-\mathbf{s}(q_z)),\qquad \left(\mathcal{M}+C\right)\ddot{q}_z= \frac{\left(1+C\right)\Gamma^2}{4\pi^2|\mathbf{q_\perp}-\mathbf{s}(q_z)|^4}(\mathbf{q_\perp}-\mathbf{s}(q_z))\cdot\frac{\partial {\bf s}}{\partial z}\biggr\rvert_{z=q_z}      \label{Eq:VortexParitlceMotion1}\\
\kappa\,   \frac{\partial\mathbf{{s}}}{\partial t}  &=& \hat{z}\times \left[ 
\frac{\Lambda \Gamma}{4\pi}\frac{\partial^2\mathbf{s}}{\partial z^2} + 
\frac{(1+C)M_0\Gamma}{\rho_\infty4\pi^2|\mathbf{q_\perp}-\mathbf{s}(z)|^4}(\mathbf{q_\perp}-\mathbf{s}(z))\delta(z-q_z)\right],\qquad\mathrm{with}\quad \kappa=\pm1,
\label{Eq:VortexParitlceMotion2}
\end{eqnarray}
where $(\hat{z}\times {\bf A})_i=\epsilon_{ij}A_j$, with $\epsilon_{ij}$ the Levi-Civita symbol. The l.h.s of Eq. (\ref{Eq:VortexParitlceMotion2}) can be straightforwardly derived following the calculations performed in references \cite{Nemirovskii2004,BustamanteNazarenko}.
Note that the r.h.s. of equation for $\ddot{q}_z$ is negligible in the limit $|\partial_z{\bf s}|\ll1$ and $|\q|\gg|{\bf s}|$.
In Eq. (\ref{Eq:VortexParitlceMotion2}) a point force is exerted by the particle giving rise to the deformation of the vortex line, while the dispersive term leads to Kelvin wave propagation. This simplified model reproduces the generation of a cusp similar to the one observed in the numerical simulations of the full GP model, as apparent in Fig. \ref{Fig:Cusp}a.
In previous works, such cusp-shaped deformations have been interpreted as the result of the vortex reconnection with the images inside the particle \cite{CaptureBerloff}. 
Such effect is not taken into account in our model and the formation of a cusp is the result of a simple action-reaction mechanism between the particle and the vortex. In addition, the curvature of the vortex filament during the trapping is not well described by the universal theoretical prediction for pure vortex reconnections obtained by Villois et al. \cite{ReconnectionGiorgio} (data not shown).

From the particle-vortex model (\ref{Eq:VortexParitlceMotion1}-\ref{Eq:VortexParitlceMotion2}) we can extract further analytical predictions. Since $|{\bf s}|\ll q_\perp$, we can set $\q_\perp-{\bf s}\approx \q_\perp$ in the model. The particle thus decouples from the vortex and just drives the forcing acting on it. We write $\mathbf{s}(z,t)$ and $\mathbf{q}(t)$ in complex variables as $s(z)=s_x(z,t)+i s_y(z,t)$ and $q(t)=|q(t)|e^{i \Omega_q(t)t}$, and linearise \eqref{Eq:VortexParitlceMotion2} for small $s$. The equation now reads
\begin{equation}
\frac{\partial \hat{s}_{k}}{\partial t}=-i\kappa\frac{\Gamma \Lambda_k}{4\pi} k^2 \hat{s}_{k} +i\frac{F_0}{|q(t)|^3}e^{-i(\Omega_q(t)+k\dot{q}_z)t}
\qquad\mathrm{with}\quad F_0=\kappa\frac{(1+C)M_0\Gamma}{4\pi^2\rho_\infty},
\label{Eq:FourierComplex}
\end{equation}
where $\hat{s}_{k}$ is the Fourier transform of $s(z)$ and $k$ a wave-vector. We have now phenomenologically included the non-local operator $\Lambda_k$ that in Fourier space reads $\Lambda_k= 2 (1-\sqrt{1-  ka_0 K_{0}(ka_0)/ K_{1}(ka_0)})/(a_0k)^2$, with $K_{n}$ the modified Bessel function of order $n$ and $a_0=1.1265\xi$. The operator $\Lambda_k$ has been defined in order to obtain the correct Kelvin wave relation dispersion $\omega_k^{\rm KW}=\kappa\Gamma\Lambda_k k^2/4\pi$ computed in references\cite{pismen1999vortices,RobertsKW} and the cut-off $a_0$ has been fixed to satisfy the known GP small-$k$ asymptotic expansion \cite{RobertsKW}.
First, let us consider the case of radial approach ($\Omega_q=0$) with vertical velocity $\dot{q}_z=0$. Integrating Eq. \eqref{Eq:FourierComplex}, it follows that the spectrum of the vortex displacement obeys the scaling $|\hat{s}_{k}|^2\sim k^{-4}$ at large scales, up to a logarithmic correction. Such scaling corresponds to the deformation of the vortex line which starts to develop already at the early times of the trapping process. We compute the spectra using the tracked vortex lines obtained from GP simulations by using the method explained in references \cite{KrstulovicKW,VilloisTracking}. They present a good agreement with theory (see Fig. \ref{Fig:Cusp}b). Finally, if $\dot{q}_z\ne0$ or $\Omega_q\ne 0$, the particle-vortex model predicts the generation of Kelvin waves when the particle is still distant. Indeed, when the particle is far from the vortex, the time dependence of $|q(t)|$ and the one of $\Omega_q(t)$ are much slower than the one of $s(t)$ and they can be treated as constants in Eq. \eqref{Eq:FourierComplex}. 
Therefore, the model \eqref{Eq:FourierComplex} predicts a linear resonance if $\omega_k^{\rm KW}=\Omega_q+k\,\dot{q}_z$. As a consequence, any motion of the particle not purely radial generates waves on the filament. In order to check this claim, we performed a GP simulation with a particle orbiting around the vortex while moving parallel to it. A movie of the simulation is available as Supplementary Information. The corresponding vortex displacement spectrum is presented in Fig. \ref{Fig:Cusp}c, where the development of a resonance is clearly visible. The resonant mode predicted by the model \eqref{Eq:FourierComplex} is $k\xi=0.24$, compatible with the position of the observed peak. In the inset a snapshot of the superfluid density shows the corresponding Kelvin waves generated on the filament. Actually, all the small-$k$ modes of the filament are growing during the first stages of particle motion. Indeed, the forcing driven by the particle produces an oscillation in time of such modes, with low frequency and high amplitude. Note that the considerations made from Eq. \eqref{Eq:FourierComplex} could be in principle formalised using a multi-time asymptotic expansion.

\section{Discussion}

We have studied the interaction of a particle and a quantum vortex in a self-consistent framework given by the particle-superfluid Hamiltonian \eqref{Eq:HGP}. The superfluid is described by the Gross-Pitaevskii equation and the particle through classical degrees of freedom. This minimal system is able to extend results obtained in more complex models \cite{CaptureBerloff,CloseParticleNewcastle} with a much lower numerical cost.
The simplicity of the model allowed us to derive the reduced Hamiltonian \eqref{Eq:HGPred} for the particle dynamics, that also includes corrections due to the vortex density profile. Similar theoretical computations can be straightforwardly performed in the case of non-local models of superfluids, that are more adequate for describing superfluid helium. In such models, the vortex density profile shows oscillations as a function of the distance to the core\cite{RobertsNonLocal,ReneuveNonLocalRec}, which could have some impact on the dynamics of small and light particles. 
In our derivation, we have neglected acoustic radiation and interaction of vortices with with sound waves. Compressibility effects of this kind might be also important for light particles and they could be included, in principle, generalising the ansatz \eqref{Eq:ansatz}. The Gross-Pitaevskii model used in this work has a very simple equation of state valid in the weak coupling limit. When the coupling is not so weak, like in superfluid helium, the equation of state can be easily modified changing the type of non-linearity of the model to account for the effect of beyond-mean-field quantum fluctuations \cite{BerloffMarcFiniteTemperature,adhikari2018vortex}. It would be interesting to study how these fluctuations modify the particle dynamics. In the same spirit, at extremely low temperatures, quantum fluctuations could excite low-amplitude Kelvin waves\cite{KrstulovicSlowdownRings}. However, we expect this effect to be negligible as it was shown in Fig.\ref{Fig:TrappingMotion}.b.

The vortex-particle interaction leads to the trapping of the particle by the vortex. The bounded state with a particle trapped inside a vortex line possesses an energy lower that the state in which the vortex and the particle are far apart. This energy gap, known as substitution energy, was first computed by Parks and Donelly \cite{DonellyParker1966SubstEnergy}. It simply corresponds to the vortex kinetic energy contained in the volume occupied by the particle. We have checked that this estimate gives the good order of magnitude for the incompressible kinetic energy difference. However, it overestimates it as it does not account for dynamical processes like the generation of Kelvin waves after the capture. The substitution energy was then used in Ref.\cite{DonellyParker1966SubstEnergy} to assess the life time of a Brownian ion inside a vortex in presence of an electric field. The model used in this Report can be trivially extended to describe a charged particle by adding an external potential, and similar considerations could be easily rephrased. 
We have observed that the vortex considerably stretches while the particle is pulled out from it by an external force (data not shown). Therefore, even at zero temperature, the energy needed to remove a particle can be much larger than the substitution energy. The release of a particle from a vortex is an interesting  problem for traditional and modern experiments with superfluids. We plan to use the model studied here to study this issue in a future work.

We have observed a non-trivial dynamics of vortex filament if a particle moves around it. The vortex dynamics has been included in the effective model \eqref{Eq:HeffV}, and we explained the motion of the vortex as the result of a mutual long-range interaction between the particle and the vortex itself. Moreover, we highlighted that long-range particle-vortex interaction is sufficient to generate Kelvin waves on the filament even if the particle never touches the vortex. It would be interesting to include such a simple interaction term in the vortex filament model, to study the effect of a large number of particles. In this regard, note that the model (\ref{Eq:GPEParticles}) can be trivially extended to include many particles, both at zero and finite temperature \cite{giuriato2018clustering}. It is then natural to use it for studying the effect of particles in a quantum turbulent regime. Indeed, it is still not clear how the dynamics of active particles modify the evolution and decay of complex tangle of quantised vortex lines. Addressing such issues is fundamental for current experiments, since particles are nowadays the main tool for tracking and visualising vortices in superfluid helium.

\appendix

\section{Numerical methods and parameters \label{App:Num}}
Equations (\ref{Eq:GPEParticles}) are solved with a standard pseudo-spectral code and a $4^{th}$ order Runge-Kutta scheme for the time stepping in a domain of size  $L$ with $N_{\rm p}$ mesh points per dimension.  We set $c=\rho_\infty=1$. 
The steady states for the particle and the vortex are prepared separately by performing imaginary time evolution of the GP equation and then they are multiplied to obtain the wished initial condition. To impose the initial flow around the particle, the initial condition is evolved for a short time ($\sim40\tau$) using GP without the particle dynamics. In Fig. \ref{Fig:Cusp}c, the target velocity in the $z$ component is reached by adding an external force $\mathbf{F}=(0,0,2\times10^{-3}M_0c^2/\xi)$ that then is switched off.

The particle potential is a smoothed hat-function $\Vp(r)=\frac{V_0}{2}(1-\tanh\left[\frac{r^2 -\eta^2}{4\Delta l^2}\right])$ and the mass displaced by the particle is measured as $M_0=\rho_\infty L^3(1-\int |\psi_\mathrm{p}|^2\,\mathrm{d}\mathbf{x}/\int |\psi_\infty|^2\,\mathrm{d}\mathbf{x})$ , where $\psi_\mathrm{p}$ is the steady state with just one particle.
Since the particle boundaries are not sharp, we measure the particle radius as $\Rp=(3M_0/4\pi\rho_\infty)^\frac{1}{3}$ for given values of the numerical parameters $\eta$ and $\Delta l$. For all the particles $V_0=20$. The parameters used are the following. For $\Rp=2.7\xi$: $N_\mathrm{p}=512$, $\eta=\xi$ and $\Delta l=0.75\xi$.
For $\Rp=7.6\xi$: $N_\mathrm{p}=256$, $\eta=2\xi$ and $\Delta l=2.5\xi$. For $\Rp=23.5\xi$:  $N_\mathrm{p}=256$,  $\eta=20\xi$ and  $\Delta l=4\xi$. Finally, for $\Rp=47\xi$:  $N_\mathrm{p}=512$,  $\eta=43\xi$ and  $\Delta l=5\xi$. Only for the last case $L=512\xi$, while for all the others $L=256\xi$.

In theoretical predictions we have used the Pad\'e approximation $\tilde{\rho}_\mathrm{v}=\bar{r}^2\left( a_1 + a_2\bar{r}^2 +a_3\bar{r}^4 \right)/\left(1+b_1\bar{r}^2+b_2\bar{r}^4 + a_3\bar{r}^6\right)$ where  $\bar{r}=|\mathbf{x}|/\xi$. The coefficients are: $a_1 =  0.340038$, $a_2 = 0.0360207$, $a_3 = 0.000985125$, $b_1 = 0.355931$, $b_2 = 0.037502$.

\section{Derivation of the reduced model for the particle trapping \label{App:Model}}
We report here the calculations leading to the reduced Hamiltonian (\ref{Eq:HGPred}).
We denote by an overbar some constants that at the leading order are independent of $\q$. The quantum energy term $H^\mathrm{GP}_\mathrm{qnt}$ contains gradients of the density and it is sub-leading when $|\q|\gg \Rp>\xi$.We thus we set $H^\mathrm{GP}_\mathrm{qnt}\approx\bar{H}^\mathrm{GP}_\mathrm{qnt}$. As discussed in the text, $1-\tilde{\rho}_\mathrm{p}(\x)$ is supported on a ball of center $\q$ and radius $\Rp$ denoted by $ \mathbb{B}(\mathbf{q},\Rp)$. 
At the leading order we also have $\tilde{\rho}_\mathrm{p}^2\approx \tilde{\rho}_\mathrm{p}$,  and $H^\mathrm{GP}_\mathrm{int}$ can be computed as
\begin{equation}
H^\mathrm{GP}_\mathrm{int}\approx \frac{g\rho_\infty}{2m^2}\int (1-(1-\tilde{\rho}_\mathrm{p}))\tilde{\rho}_{\rm v}^2\,\mathrm{d}\mathbf{x}=
\bar{H}^\mathrm{GP}_\mathrm{int}-\frac{g\rho_\infty^2}{2m^2}\int\left(1-\tilde{\rho}_\mathrm{p}\right)\tilde{\rho}_\mathrm{v}^2\,\mathrm{d}\mathbf{x}\approx
\bar{H}^\mathrm{GP}_\mathrm{int}-\frac{g\rho_\infty^2}{2m^2}\int\limits_{\mathbb{B}(\mathbf{q},\Rp)}\tilde{\rho}_\mathrm{v}^2\,\mathrm{d}\mathbf{x}\approx \bar{H}^\mathrm{GP}_\mathrm{int}-\frac{1}{2}M_0c^2\tilde{\rho}_\mathrm{v}^2(q_\perp).
\label{Eq:Hint_M}
\end{equation} 
where we have treated $\tilde{\rho}_\mathrm{v}$ as constant inside $ \mathbb{B}(\mathbf{q},\Rp)$, recognised the displaced superfluid mass $M_0=\rho_\infty\int_{\mathbb{B}}\mathrm{d}\mathbf{x}$ and used the definition of the speed of sound. A similar calculation can be performed for the the term $H^\mathrm{GP}_\mathrm{p}$:
\begin{equation}
H^\mathrm{GP}_\mathrm{p}\approx
\bar{H}^\mathrm{GP}_\mathrm{p} - \frac{\mu\rho_\infty}{m}\int\left(1-\tilde{\rho}_\mathrm{p}\right)\tilde{\rho}_\mathrm{v}\,\mathrm{d}\mathbf{x}\approx
\bar{H}^\mathrm{GP}_\mathrm{p}-\frac{\mu\rho_\infty}{m}\int_{\mathbb{B}(\mathbf{q},\Rp)}\tilde{\rho}_\mathrm{v}\,\mathrm{d}\mathbf{x}\approx\bar{H}^\mathrm{GP}_\mathrm{p}+M_0c^2\tilde{\rho}_\mathrm{v}(q_\perp).
\label{Eq:Hp_M}
\end{equation} 
In the first equality we used the Thomas-Fermi approximation (\ref{Eq:RhoPart}) and again $\tilde{\rho}_\mathrm{p}^2\approx \tilde{\rho}_\mathrm{p}$. 

To compute the hydrodynamic term $H^\mathrm{GP}_\mathrm{hydro}$ we write $\mathbf{v}_\mathrm{s}^2=\mathbf{v}_{\rm v}^2+\mathbf{v}_\mathrm{p}\cdot(2\mathbf{v}_{\rm v}+\mathbf{v}_\mathrm{p}) + \mathbf{v}_\mathrm{BC}\cdot(2\mathbf{v}_{\rm v}+2\mathbf{v}_\mathrm{p}+\mathbf{v}_\mathrm{BC})$ and consider the contribution of the three addends differently. Firstly we have
\begin{equation}
H^\mathrm{GP(v)}_\mathrm{hydro}=\frac{1}{2}\int\rho\mathbf{v}_\mathrm{v}^2\,\mathrm{d}\mathbf{x}
=
\bar{H}^\mathrm{GP}_\mathrm{hydro}-\frac{\rho_\infty}{2}\int\left(1-\tilde{\rho}_\mathrm{p}\right)\tilde{\rho}_\mathrm{v}\mathbf{v}_{\rm v}^2\,\mathrm{d}\approx \bar{H}^\mathrm{GP}_\mathrm{hydro}-\frac{M_0 \tilde{\rho}_\mathrm{v}(q_\perp)}{2}\mathbf{v}_{\rm v}(q_\perp)^2
= \bar{H}^\mathrm{GP}_\mathrm{hydro}-\frac{\Gamma^2M_0\tilde{\rho}_\mathrm{v}(q_\perp)}{8\pi^2q_\perp^2}.
\label{Eq:Hhydrov_M}
\end{equation}
The second term is computed integrating by parts, using the incompressibility of the flows and neglecting the gradients of $\rho_{\rm v}$:
\begin{equation}
H^\mathrm{GP(p)}_\mathrm{hydro}=\frac{\rho_\infty}{2}\int\tilde{\rho}_{\rm p}\tilde{\rho}_{\rm v} \nabla\phi_\mathrm{p}\cdot(2\mathbf{v}_{\rm v}+\mathbf{v}_\mathrm{p})  \,\mathrm{d}\mathbf{x}
\approx \frac{\rho_\infty}{2}\int\limits_{\mathbb{B}^{\rm c}(\mathbf{q},\Rp)}\tilde{\rho}_{\rm v} \nabla\phi_\mathrm{p}\cdot(2\mathbf{v}_{\rm v}+\mathbf{v}_\mathrm{p}) \mathrm{d}\mathbf{x}
\approx -\frac{\rho_\infty\Rp^2}{2}\oint\limits_{\partial\mathbb{B}(\mathbf{q},\Rp)}\tilde{\rho}_{\rm v} \phi_\mathrm{p}(2\mathbf{v}_{\rm v}+\mathbf{v}_\mathrm{p})\cdot \hat{n}\mathrm{d}{\Omega}, 
\label{Eq:Hhydrovp_M}
\end{equation}
where $\mathbb{B}^{\rm c}(\mathbf{q},\Rp)$ is the complement of $\mathbb{B}(\mathbf{q},\Rp)$ and $\partial \mathbb{B}(\mathbf{q},\Rp)$ its boundary.
For  $\epsilon\ll1$ (see Eq.\eqref{Eq:epsilon}), in the last integral $\rho_{\rm v}$ and $\mathbf{v}_{\rm v}$ can be evaluated at $\q$. The angular integral is then computed exactly and it gives $H^\mathrm{GP(vp)}_\mathrm{hydro}=\frac{M_0 \tilde{\rho}_{\rm v} (\q)}{2} C (\dot{\q}^2-\mathbf{v}_{\rm v}^2)$, with $C=1/2$.

The velocity ${\bf v}_{\rm BC}$ is obtained at order $\epsilon^2$ from the potential $\phi_\mathrm{BC}=\frac{1}{2}e_{ij}x'_ix'_j$, where $e_{ij}(\q)=(\partial_i v_{\rm v}^j+\partial_j v_{\rm v}^i)/2$ is the strain rate tensor evaluated at the particle position. 
Following the same procedure of \eqref{Eq:Hhydrovp_M}, we see that the only non-zero contribution is given by
\begin{equation}
H^\mathrm{GP(BC)}_\mathrm{hydro}= -\frac{\rho_\infty\Rp^2}{2}\oint\limits_{\partial\mathbb{B}(\mathbf{q},\Rp)}\tilde{\rho}_{\rm v} \phi_\mathrm{BC}\mathbf{v}_\mathrm{BC}\cdot \hat{n}\mathrm{d}{\Omega}= -\frac{\rho_\infty\Rp^5}{4}\tilde{\rho}_{\rm v}\,e_{ij}e_{rs}\oint \hat{n}_i\hat{n}_j\hat{n}_r\hat{n}_s \mathrm{d}{\Omega}
=-\frac{M_0\tilde{\rho}_{\rm v}\Rp^2}{10}(e_{ij})^2=-\frac{M_0\Gamma^2\tilde{ \rho}_{\rm v}\Rp^2}{20\pi^2 q_\perp^4}.
\label{Eq:HhydroHOT_M}
\end{equation}
We have used $\oint \hat{n}_i\hat{n}_j\hat{n}_r\hat{n}_s \mathrm{d}{\Omega}=(\delta_{ij}\delta_{rs}+\delta_{ir}\delta_{js}+\delta_{is}\delta_{ir})\Omega_d/d(d+2)$,  where $\Omega_d$ is the surface of the unit sphere in $d$ dimensions. The total hydrodynamic kinetic term is the just given by $H^\mathrm{GP}_\mathrm{hydro}=H^\mathrm{GP(v)}_\mathrm{hydro}+H^\mathrm{GP(p)}_\mathrm{hydro}+H^\mathrm{GP(BC)}_\mathrm{hydro}$. 

Note that at the order $\xi^4/q_\perp^4$ for large vortex-particle separations, the reduced Hamilonian (\ref{Eq:HGPred}) becomes
\begin{equation}
H_\mathrm{red}[\mathbf{q},\mathbf{p}]=
\bar{H}^\mathrm{GP}+\frac{\mathbf{p}^2}{2M_\mathrm{eff}}+M_0c^2\left[\frac{1}{2}-\frac{(1+C)\xi^2}{q_\perp^2}+\frac{(10C+5-4(\Rp/\xi)^2)}{10}\frac{\xi^4}{q_\perp^4}
\right],
\label{Eq:HGPredFullFar}
\end{equation}
where the term proportional to $\xi^4/q_\perp^{4}$ turns out to be repulsive if $\Rp^2<\frac{5}{4}(1+2C)\xi^2$.

\section{Vortex deformation \label{App:Vortex}}

We can give a rough derivation of the effective vortex-particle Hamiltonian \eqref{Eq:HeffV} by assuming a small deformation of the vortex line.
Similarly to the previous calculations, we neglect the gradients of $\tilde{\rho}_{\rm p}$. At the leading order, the terms $H^\mathrm{GP}_\mathrm{hydro}$ and  $H^\mathrm{GP}_\mathrm{qnt}$ are the only contributing to $H_\mathrm{v-p}$. For distant particles, the only component of the superfluid velocity modified by the vortex displacement is $\mathbf{v}_\mathrm{v}$. The new contributions to the energy coming from the products of $\mathbf{v}_{\rm v}$ and the other velocity fields vanish after angular integration. 
We thus need to compute the term $\mathbf{v}_{\rm v}(|\x - {\bf R}(z)|)^2=\nabla_\perp\phi_{\rm v}(\x_\perp - {\bf s}(z))^2 + (\nabla_\perp\phi_{\rm v}(|\x_\perp - {\bf s}(z)|)\cdot \partial_z {\bf s}(z) )^2 $ where now $\x_\perp=(x,y)$ and $\nabla_\perp=(\partial_x,\partial_y)$. The first term leads directly to the potential (\ref{Eq:Hhydro}) evaluated at $\q_\perp - {\bf s}(q_z)$. The second term is treated by using the fact that $1-\tilde{\rho}_{\rm p}$ is practically supported on $\mathbb{B}(\mathbf{q},\Rp)$ and $\tilde{\rho}_{\rm p}=1-(1-\tilde{\rho}_{\rm p})$. We need to compute
\begin{eqnarray}
\nonumber H^\mathrm{GP}_{\rm LIA}&\approx& \frac{\rho_\infty}{2}\int\tilde{\rho}_{\rm v}(|\x_\perp - {\bf s}(z)|)(\nabla_\perp\phi_{\rm v}(|\x_\perp - {\bf s}(z)|)\cdot \partial_z {\bf s}(z) )^2\,\mathrm{d}\mathbf{x}
-\frac{\rho_\infty}{2}\int\limits_{\mathbb{B}(\mathbf{q},\Rp)}\tilde{\rho}_{\rm v}(|\x_\perp - {\bf s}(z)|)(\nabla_\perp\phi_{\rm v}(|\x_\perp - {\bf s}(z)|)\cdot \partial_z {\bf s}(z) )^2\,\mathrm{d}\mathbf{x}\\
&\approx&\frac{\rho_\infty}{2}\int_0^L 
\partial_z { s_i}(z)\partial_z { s_j}(z)\left(\int \tilde{\rho}_{\rm v}(x_\perp) v_{\rm v}^i(x_\perp)v_{\rm v}^j  \mathrm{d}{\bf x}_\perp\right)\mathrm{d}z+\frac{M_0}{2}\tilde{\rho}_{\rm v}(q_\perp)(\mathbf{v}_{\rm v}(q_\perp)\cdot \partial_z {\bf R}(z))^2  \label{Eq:HLIAderivation}\\
&\approx&\frac{\rho_\infty}{4}\int_0^L 
(\partial_z {\bf s}(z) )^2\left(\int \tilde{\rho}_{\rm v}(x_\perp) \mathbf{v}_{\rm v}(x_\perp)^2  \mathrm{d}{\bf x}_\perp\right)\mathrm{d}z= \frac{\rho_\infty \Gamma^2\Lambda_{\rm hydro}}{8\pi}\int_0^L
(\partial_z {\bf s}(z) )^2 \mathrm{d}z,
\label{Eq:HGPredFullFar}
\end{eqnarray}
where we have neglected the second term in \eqref{Eq:HLIAderivation} as it subdominant. The constant $\Lambda_{\rm hydro}$ is given by the radial integral 
\begin{equation}
\Lambda_{\rm hydro}=\int_0^{L_{\rm d}} \frac{\tilde{\rho}_{\rm v}(r)}{r}\mathrm{d}r=\int_0^\xi \frac{\tilde{\rho}_{\rm v}(r)}{r}\mathrm{d}r+\int_\xi^\infty \frac{\tilde{\rho}_{\rm v}(r)-1}{r}\mathrm{d}r+\log{\frac{L_{\rm d}}{\xi}}+O((L_{\rm d}/\xi)^2)=\log{\frac{L_{\rm d}}{\xi}}-0.3978+O((L_{\rm d}/\xi)^2),
\end{equation}
where $L_{\rm d}$ is a cut-off of the order of the inter-vortex distance.
The last integral has been performed numerically using $\tilde{\rho}_{\rm v}$ obtained by imaginary time evolution of the GP equation in a infinite domain. The quantum energy term $H^\mathrm{GP}_\mathrm{qnt}$ gives a contribution equal to \eqref{Eq:HGPredFullFar} but with the constant $\Lambda_{\rm hydro}$ replaced by $\Lambda_{\rm qnt}=\frac{1}{4}\int_0^\infty \left(  \frac{\mathrm{d} \tilde{\rho}_{\rm v}}{\mathrm{d}r} \right)^2 \frac{r}{ \tilde{\rho}_{\rm v}}\mathrm{d}r=0.2854$. Finally, in Eq. \eqref{Eq:HeffV} the constant is $\Lambda=\Lambda_{\rm hydro}+\Lambda_{\rm qnt}$. This oversimplified derivation does not recover the full dispersion relation of Kelvin waves as it neglects non-linear interactions and the 3D modifications of ${\bf v}_{\rm v}$ due to vortex deformations. For a more accurate discussion see references \cite{RobertsKW,pismen1999vortices,LaurieKWPRB}.

\section*{Acknowledgements}
We acknowledge useful scientific discussions with Marc Brachet, Gustavo During, Davide Proment, Vishwanath Shukla and Sergey Nazarenko. The authors were supported by Agence Nationale de la Recherche through the project GIANTE ANR-18-CE30-0020-01. Computations were carried out on the M\'esocentre SIGAMM hosted at the Observatoire de la C\^ote d'Azur and the French HPC Cluster OCCIGEN through the GENCI allocation A0042A10385.

\bibliographystyle{apsrev4-1}
%merlin.mbs apsrev4-1.bst 2010-07-25 4.21a (PWD, AO, DPC) hacked
%Control: key (0)
%Control: author (8) initials jnrlst
%Control: editor formatted (1) identically to author
%Control: production of article title (-1) disabled
%Control: page (0) single
%Control: year (1) truncated
%Control: production of eprint (0) enabled
%
%\bibliography{../RefsVortexParticles}

\end{document}